\documentclass[epj,twocolumn]{webofc}
\usepackage[varg]{txfonts}   
\woctitle{Flavour changing and conserving processes}
\newcommand{\gv}{\mbox{GeV}}

\newcommand{\ppm}{\pi^+ \pi^-}
\newcommand{\eepp}{e^+e^- \rightarrow \pi^+\pi^-}
\newcommand{\x}{\phantom{x}}
\newcommand{\bl}{\phantom{-}}

\newcommand{\mbo}[1]{$#1$ }
\newcommand{\epm}{e^+e^- }
\newcommand{\epem}{e^+e^- }
\newcommand{\power}[1]{\times 10^{#1} }
\newcommand{\I}{\rm i }

\newcommand{\Impa}{\rm Im }
\newcommand{\Repa}{\rm Re }
\newcommand{\semis}{\;;\;\; }
\newcommand{\amu}{a_\mu }
\newcommand{\amuh}{a_\mu^{\rm had} }
\newcommand{\bea}{\begin{eqnarray}}
\newcommand{\eea}{\end{eqnarray}}
\newcommand{\epo}{\;. }
\newcommand{\cA}{{\cal A} }

\newcommand{\cL}{{\cal L} }

\newcommand{\nn}{\nonumber}
\newcommand{\bary}{\begin{array}}
\newcommand{\eary}{\end{array}}

\newcommand{\ttc}[1]{\multicolumn{2}{c}{#1}}

\begin{document}
\title{Leading-order hadronic contribution to the electron and muon $g-2$}
%
%

\author{Fred Jegerlehner\inst{1,2}\fnsep\thanks{\email{fjeger@physik.hu-berlin.de}} 
}

\institute{
Deutsches  Elektronen--Synchrotron (DESY), Platanenallee 6, D--15738 Zeuthen, Germany
\and
Humboldt--Universit\"at zu Berlin, Institut f\"ur Physik, Newtonstrasse 15, D--12489 Berlin,
Germany
          }

\abstract{%
I present a new data driven update of the hadronic vacuum polarization
effects for the muon and the electron $g-2$. For the leading order
contributions I find $\amu^{\mathrm{had}(1)}=(686.99\pm
4.21)[687.19\pm 3.48]\times 10^{-10}$ based on $\epm$data [incl. $\tau$
data], $\amu^{\mathrm{had}(2)}= (-9.934\pm 0.091) \times 10^{-10}$ (NLO)
and $\amu^{\mathrm{had}(3)}= (1.226\pm 0.012) \times 10^{-10}$ (NNLO)
for the muon, and $a_e^{\mathrm{had}(1)}=(184.64\pm 1.21)\times 10^{-14}$
(LO), $a_e^{\mathrm{had}(2)}=(-22.10\pm 0.14)\times 10^{-14}$ (NLO) and
$a_e^{\mathrm{had}(3)}=(2.79\pm 0.02)\times 10^{-14}$ (NNLO)
for the electron.  A problem with vacuum polarization undressing of
cross-sections (time-like region) is addressed. I also add a comment
on properly including axial mesons in the hadronic light-by-light
scattering contribution.  My estimate here reads $\amu[a_1,f_1',f_1]
\sim ({ 7.51 \pm 2.71}) \power{-11}\,.$ With these  updates
$a_\mu^{\rm exp}-a_\mu^{\rm the}=(32.73\pm 8.18)\times 10^{-10}$ a 4.0
$\sigma$ deviation, while $a_e^{\rm exp}-a_e^{\rm the}=(-1.10\pm
0.82)\times 10^{-12}$ shows no significant deviation.}
%
\onecolumn
\thispagestyle{empty}
\begin{flushright}
\large
DESY~15-220,~~HU-EP-15/53\\
November 2015
\end{flushright}

\vfill

\begin{center}
{\huge\bf
Leading-order hadronic contribution to the electron and muon $g-2$}\\[1.5cm]
{\large\bf Fred Jegerlehner}\\[1cm]
\textit{Deutsches  Elektronen--Synchrotron (DESY), Platanenallee 6,\\ D--15738 Zeuthen, Germany\\
Humboldt--Universit\"at zu Berlin, Institut f\"ur Physik, Newtonstrasse 15,\\ D--12489 Berlin,
Germany}

\vfill

{\large\bf Abstract}\\[3mm]

\begin{minipage}{0.8\textwidth}
I present a new data driven update of the hadronic vacuum polarization
effects for the muon and the electron $g-2$. For the leading order
contributions I find $\amu^{\mathrm{had}(1)}=(686.99\pm
4.21)[687.19\pm 3.48]\times 10^{-10}$ based on $\epm$data [incl. $\tau$
data], $\amu^{\mathrm{had}(2)}= (-9.934\pm 0.091) \times 10^{-10}$ (NLO)
and $\amu^{\mathrm{had}(3)}= (1.226\pm 0.012) \times 10^{-10}$ (NNLO)
for the muon, and $a_e^{\mathrm{had}(1)}=(184.64\pm 1.21)\times 10^{-14}$
(LO), $a_e^{\mathrm{had}(2)}=(-22.10\pm 0.14)\times 10^{-14}$ (NLO) and
$a_e^{\mathrm{had}(3)}=(2.79\pm 0.02)\times 10^{-14}$ (NNLO)
for the electron.  A problem with vacuum polarization undressing of
cross-sections (time-like region) is addressed. I also add a comment
on properly including axial mesons in the hadronic light-by-light
scattering contribution.  My estimate here reads $\amu[a_1,f_1',f_1]
\sim ({ 7.51 \pm 2.71}) \power{-11}\,.$ With these  updates
$a_\mu^{\rm exp}-a_\mu^{\rm the}=(32.73\pm 8.15)\times 10^{-10}$ a 4.0
$\sigma$ deviation, while $a_e^{\rm exp}-a_e^{\rm the}=(-1.10\pm
0.82)\times 10^{-12}$ shows no significant deviation.
\end{minipage}
\end{center}
\vfill
\noindent\rule{8cm}{0.5pt}\\
$^*$ Invited talk  FCCP2015 -
 Workshop on ``Flavour changing and conserving processes,'' 10-12 September
2015,
Anacapri, Capri Island, Italy.
\setcounter{page}{0}
\newpage
\twocolumn
%
\maketitle
\section{Introduction: hadronic effects in $g-2$.}
\label{intro}
A well known general problem in electroweak precision physics are the
higher order contributions from hadrons (quark loops) at low energy
scales.  While leptons primarily exhibit the fairly weak
electromagnetic interaction, which can be treated in perturbation
theory, the quarks are strongly interacting via confined gluons where
any perturbative treatment breaks down. Considering the lepton
anomalous magnetic moments one distinguishes three types of
non-perturbative corrections:
\textbf{(a)} Hadronic Vacuum Polarization (HVP) of order
    $O(\alpha^2),O(\alpha^3),O(\alpha^4)$;
\textbf{(b)} Hadronic Light-by-Light (HLbL) scattering at $O(\alpha^3)$;
\textbf{(c)} hadronic effects at $O(\alpha G_F m_\mu^2)$ in 2-loop electroweak (EW)  
corrections, in all cases quark-loops appear as hadronic
``blobs''. The hadronic contributions are limiting the precision of
the predictions.

Evaluation of non-perturbative effects is possible by using
experimental data in conjunction with Dispersion Relations (DR), by
low energy effective modeling via a Resonance Lagrangian Approach
(RLA) ( Vector Meson Dominance (VMD) implemented in accord with chiral
structure of QCD)~\cite{HKS95,BPP1995,Bijnens15}, like the Hidden
Local Symmetry (HLS) or the Extended Nambu Jona-Lasinio (ENJL) models,
or by lattice QCD.  Specifically:
\textbf{(a)} HVP via a dispersion integral over
\mbo{\epem \to \mathrm{hadrons}} data
(1 independent amplitude to be determined by one specific data channel)
(see e.g. ~\cite{EJ95}) as elaborated below, by the HLS effective Lagrangian 
approach~\cite{Benayoun:2011mm,Benayoun:2012wc,Benayoun:2015gxa,Jegerlehner:2013sja,Maurice15}, 
or by lattice QCD~\cite{Boyle:2011hu,Feng:2013xsa,Aubin:2013daa,Francis:2014dta,Malak:2015sla,Petschlies15};
\textbf{(b)} HLbL via a RLA together with operator product expansion
(OPE) methods~\cite{KnechtNyffeler01,MV2004,Knecht15,Nyffeler15}, by a dispersive approach using \mbo{\gamma\gamma \to
\mathrm{hadrons}} data (28 independent
amplitudes to be determined by as many independent data sets in
principle)~\cite{PaukVanderhaeghen2013,Colangelo:2014pva,Procura15} or
by lattice QCD~\cite{Blum:2014oka,Lehner15};
\textbf{(c)} EW quark-triangle diagrams are well under control,
because the possible large corrections are related to the
Adler-Bell-Jackiw (ABJ) anomaly which is perturbative and
non-perturbative at the same time. Since VVV = 0 by the Furry theorem,
only VVA (of \mbo{\gamma \gamma Z}-vertex, V=vector, A=axialvector)
contributes. In fact leading effects are of short distance type ($M_Z$
mass scale) and cancel against
lepton-triangle loops (anomaly cancellation)~\cite{KPPdeR02,CMV03}.
 
\section{Leading-order $a_\mu^{\rm had}$ via
\mbo{\sigma(\epem \to \mathrm{hadrons})}}
\label{sec-1}
\begin{figure}[h]
\centering
\includegraphics[width=2.5cm]{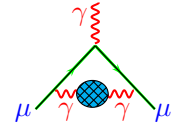}
\caption{Leading diagram exhibiting a hadronic ``blob''.}
\label{fig:amudiag} 
\end{figure}

\begin{figure}
\centering
\includegraphics[width=7cm]{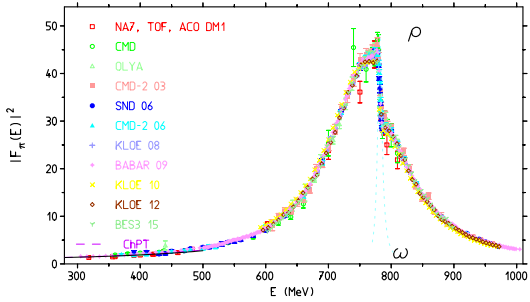}
\caption{The pion form factor $|F_\pi(s)|^2=4\,R_{\pi\pi}/\beta_\pi^3
$ ($\beta_\pi=\sqrt(1-4m_\pi^2/s)$) dominated by the $\rho$ resonance
peak. Data include measurements from
Novosibirsk (NSK) ~\cite{CMD203,CMD206,SND06}, Frascati
(KLOE)~\cite{KLOE08,KLOE10,KLOE12}, SLAC (BaBar)~\cite{BABARpipi} and
Beijing (BESIII)~\cite{BESIII}.}
\label{fig:ffpp} 
\end{figure}

\begin{figure}
\centering
\includegraphics[width=7cm]{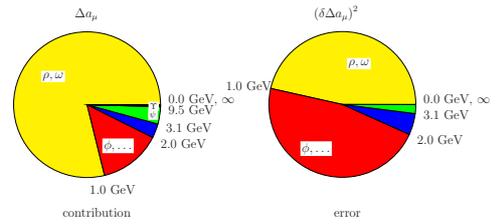}
\caption{Muon $g-2$: distribution of contributions and error squares from 
different energy ranges.}
\label{fig:gmdist} 
\end{figure}

\begin{figure}
\centering
\includegraphics[width=7cm]{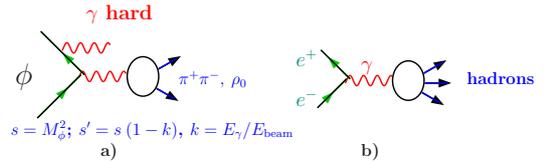}
\caption{a) Initial state radiation (ISR), b) Standard energy scan.}
\label{fig:RR}
\end{figure}

\begin{figure}
\centering
\includegraphics[width=5.5cm]{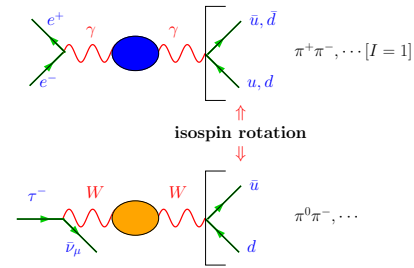}
\caption{$\tau$-decay data may be combined with I=1 part of $\epm$
annihilation data after isospin rotation $[\pi^-\pi^0] \Leftrightarrow [\pi^-\pi^+]$ and applying isospin breaking
(IB) corrections (e.m. effects, phase space, isospin breaking in
masses, widths, $\rho^0-\omega$ mixing etc.).}
\label{fig:tauvsee}
\end{figure}

The leading \textit{non-perturbative} hadronic contribution to
$\amuh$, represented by the diagram figure~\ref{fig:amudiag}, 
can be obtained in terms of undressed experimental cross-sections data
\bea
R_\gamma(s) \equiv { \sigma^{(0)}(e^+e^-\rightarrow \gamma^*
\rightarrow {\rm hadrons})}/{ \frac{4\pi \alpha^2}{3s}}\,,
\label{Rfun}
\eea 
$s=E_{\rm cm}^2$, $E_{\rm cm}$ the center of mass energy, and the DR:
{\small
\bea
{\amuh} &=& \left(\frac{\alpha m_\mu}{3\pi}
\right)^2 \bigg(\;\;\;
\int\limits_{4 m_\pi^2}^{E^2_{\rm cut}}ds\,
\frac{{ R^{\mathrm{data}}_\gamma(s)}\;\hat{K}(s)}{s^2}
\nonumber \\ && \hspace*{1cm} + \int\limits_{E^2_{\rm cut}}^{\infty}ds\,
\frac{{ R^{\mathrm{pQCD}}_\gamma(s)}\;\hat{K}(s)}{s^2}\,\,
\bigg)\epo
\label{amuint}
\eea
} The kernel $\hat{K}(s)$ is an analytically known monotonically
increasing function, raising from about 0.64 at the two pion threshold
$4 m_\pi^2$ to 1 as $s \to \infty$.  This integral is well defined
due to the asymptotic freedom of QCD, which allows for a
perturbative QCD (pQCD) evaluation of the high energy
contributions. Because of the $1/s^2$ weight, the dominant
contribution comes from the lowest lying hadronic resonance, the
$\rho$ meson (see figure~\ref{fig:ffpp}).  As low energy contributions
are enhanced, about $\sim 75\% $ come from the region
$2m_\pi\!<\!\!\!\sqrt{s}\!<\!\!1\,\gv$ dominated by the $\pi^+\pi^-$
channel. Experimental errors imply theoretical uncertainties, the main
issue for the muon $g-2$.
Typically, results are collected from different resonances and regions
as presented in table~\ref{tab:amures}. Statistical errors (stat) are
summed in quadrature, systematic (syst) ones are taken into account
linearly (100\% correlated) within the different contributions of the list, and
summed quadratically from the different regions and resonances. From
5.2 GeV to 9.46 GeV and above 13 GeV pQCD is used.
Relative (rel) and  absolute (abs) errors are also shown.
\begin{table}[t]
\label{tab:amures}
\centering
{\scriptsize
\caption{Results for \mbo{a_\mu^{\mathrm{had(1)}}} (in units $\power{-10}$).}
\begin{tabular}{|cc||r|r|}
\hline
\!\!\!final state \!\!\! &  range (GeV) & \mbo{a_\mu^{\mathrm{had(1)}}} (stat) (syst) [tot] & rel[abs]\% \\
\hline
 $\rho  $    & ( 0.28, 1.05) &    505.96 ( 0.77) ( 2.47)[ 2.59]&   0.5 [37.8] \\
 $\omega $    & ( 0.42, 0.81) &     35.23 ( 0.42) ( 0.95)[ 1.04]&  3.0
 [\x6.1] \\
 $\phi   $    & ( 1.00, 1.04) &     34.31 ( 0.48) ( 0.79)[ 0.92]&  2.7
 [\x4.8] \\
 $J/\psi $    &             &      8.94 ( 0.42) ( 0.41)[ 0.59]&    6.6
 [\x1.9] \\
 $\Upsilon$   &             &      0.11 ( 0.00) ( 0.01)[ 0.01]&    6.8
 [\x0.0] \\
   had        & ( 1.05, 2.00) &     60.45 ( 0.21) ( 2.80)[ 2.80]&  4.6 [44.4] \\
   had        & ( 2.00, 3.10) &     21.63 ( 0.12) ( 0.92)[ 0.93]&  4.3
   [\x 4.8] \\
   had        & ( 3.10, 3.60) &      3.77 ( 0.03) ( 0.10)[ 0.10]&  2.8
   [\x 0.1] \\
   had        & ( 3.60, 5.20) &      7.50 ( 0.04) ( 0.01)[ 0.04]&  0.3
   [\x 0.0] \\
  pQCD        & ( 5.20, 9.46) &      6.27 ( 0.00) ( 0.01)[ 0.01]&  0.0
   [\x 0.0] \\
   had        & ( 9.46,13.00) &      1.28 ( 0.01) ( 0.07)[ 0.07]&  5.4
   [\x 0.0] \\
  pQCD        &(13.0,$\infty$)&      1.53 ( 0.00) ( 0.00)[ 0.00]&  0.0
  [\x 0.0] \\
\hline
  data        & ( 0.28,13.00) &    679.19 ( 1.12) ( 4.06)[ 4.21]&  0.6  [100.] \\
  total       &             &    686.99 ( 1.12) ( 4.06)[ 4.21]&  0.6 [100.] \\
\hline
\end{tabular}
}
\end{table}
The distribution of contributions and errors are illustrated in the
pie chart figure~\ref{fig:gmdist}.
As a result we find
\bea
\amu^{\mathrm{had}(1)}=(686.99\pm 4.21)[687.19\pm3.48]\times 10^{-10}
\label{LOHVPres}
\eea
based on $\epm$--data [incl. $\tau$-decay spectra~\cite{JS11}].  In
the last 15 years $\epm$ cross-section measurements have dramatically
improved, from energy scans~\cite{CMD203,CMD206,SND06} at Novosibirsk
(NSK) and later, using the radiative return mechanism, measurements
via initial state radiation (ISR) at meson factories (see
figure~\ref{fig:RR})~\cite{KLOE08,KLOE10,KLOE12,BABARpipi,BESIII}.
A third possibility to enhance experimental information useful to
improve HVP estimates are \mbo{\tau}--decay spectra $\tau \to \bar{\nu}_\tau
\pi^0\pi^-,\cdots$, supplied by isospin breaking
effects~\cite{ADH98,DEHZ03,GJ04,Davier:2009ag,JS11,Benayoun:2011mm,Benayoun:2012wc,Benayoun:2015gxa,ZhiqingZhang15}. 
In the conserved vector current (CVC) limit $\tau$ spectra
should be identical to the isovector part $I=1$ of the $\epm$ spectra,
as illustrated in figure~\ref{fig:tauvsee}.
Including the \mbo{I=1} \mbo{\tau \to \pi\pi \nu_\tau} data available
from~\cite{ALEPH,AlephCorr,OPAL,CLEO,Belle} in the range [0.63-0.96]
GeV one obtains ~\cite{JS11}:
\bea
\!\!\!\!\amuh[ee\to\pi\pi]\!\! &=&\!\! {353.82}(0.88)(2.17)[{ 2.34}]\power{-10}\nn \\
\!\!\!\!\amuh[\tau\to\pi\pi\nu]\!\! &=&\!\! {354.25}(1.24)(0.61)[{
1.38}]\power{-10}\nn \\
\!\!\!\!\amuh[\x ee + \tau\x]\!\! &=&\!\! {354.14}(0.82)(0.86)[{ 1.19}]\power{-10}\,,\nn
\eea
which improves the LO HVP as given in (\ref{LOHVPres}).
We briefly summarize recent progress in data collection as follows.
\subsection{Data}
\begin{figure}
\centering
\includegraphics[width=7cm]{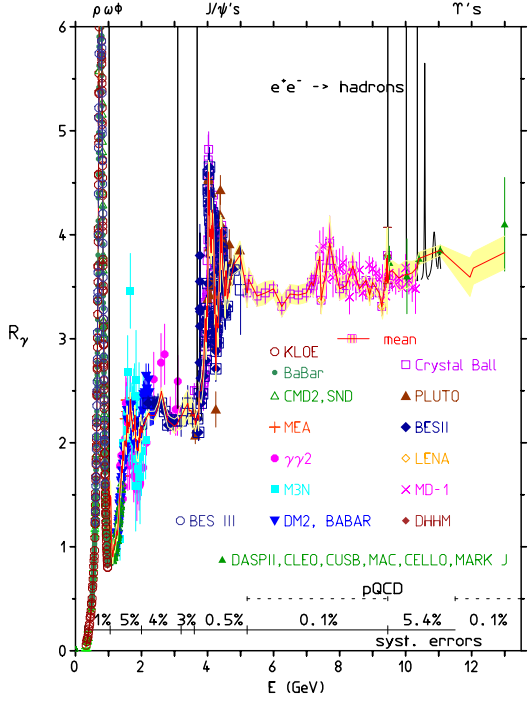}
\caption{My 2015 compilation of $R_\gamma$ as a function of energy $E$.}
\label{fig:Rcompi} 
\end{figure}

\begin{figure}[h]
\centering
\includegraphics[width=7cm]{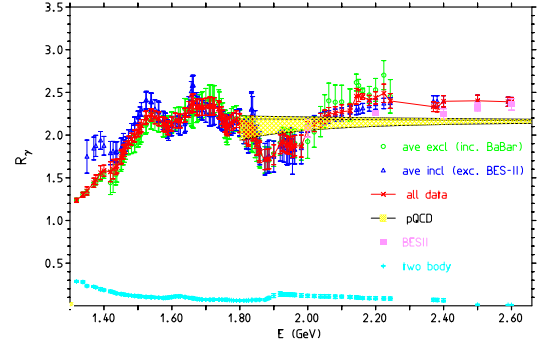}
\caption{$e^+e^-$ annihilation data in the 1.4 to 2.6 GeV
region. Summed up exclusive (excl) channel data are shown together
with old inclusive data (incl). Two-body channels represent a small
fraction of $R_\gamma$ only. Above 2 GeV good quality inclusive BES-II 
data~\cite{BES02}
provide a fairly well determined $R_\gamma(s)$.}
\label{fig:exclvsincl} 
\end{figure}

As I mentioned the most important data are the $\pi\pi$ production
data in the range up to 1 GeV. New experimental input for HVP comes
from BESIII~\cite{BESIII}. Still the most precise ISR measurements
from KLOE and BaBar are in conflict and the new, although still
somewhat less precise, ISR data from BESIII help to clarify this
tension. The BESIII result for $a_\mu^{\pi\pi,
\mathrm{LO}}(0.6-0.9~\gv)$ is found to be in good agreement with all
KLOE values, while a 1.7 $\sigma$ lower value is observed with respect
to the BaBar result. Other data recently collected,
and published up to the end of 2014, include the $e^+e^-\to
3(\pi^+\pi^-)$ data from CMD--3
\cite{Akhmetshin:2013xc}, the $e^+e^- \to \omega\pi^0
\to \pi^0\pi^0\gamma$ from SND \cite{Achasov:2013btb} and several data
sets collected by BaBar in the ISR mode\footnote{Including the $p
\bar{p}$,~$K^+K^-$,~$K_LK_S,\:K_LK_S\pi^+\pi^-$,~$K_SK_S\pi^+\pi^-,K_S
K_S K^+K^-$ final
states.}~\cite{Lees:2013ebn,Lees:2013gzt,Lees:2014xsh,Davier:2015bka}. These
data samples highly increase the available statistics for the
annihilation channels opened above 1 GeV and lead to significant
improvements. Recent/preliminary results also included are \mbo{\epm
\to \pi^+\pi^-\pi^0} from Belle, \mbo{\epm \to K^+K^-} from CMD-3,
\mbo{\epm \to K^+K^-} from SND.
The resulting data sample is collected in figure~\ref{fig:Rcompi},
which has indicated the overall precision of the different ranges as
well as the pQCD ranges, where data are replaced by pQCD results.
Still one of the main issue in HVP is \mbo{R_\gamma(s)} in the region 1.2 to
2.4 GeV, which actually has been improved dramatically by the
exclusive channel measurements by BaBar in the last decade. The most
important 20 out of more than 30 channels are measured, many known at
the 10 to 15\% level. The exclusive channel therefore has a much
better quality than the very old inclusive data from Frascati (see
figure~\ref{fig:exclvsincl}).

\subsection{NLO and NNLO HVP effects updated}
\begin{figure}
\centering
\includegraphics[width=6cm]{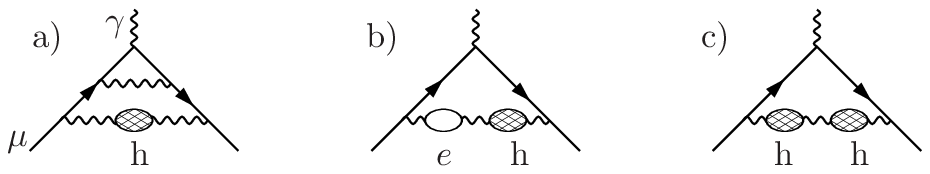}\\[3mm]
\includegraphics[width=8cm]{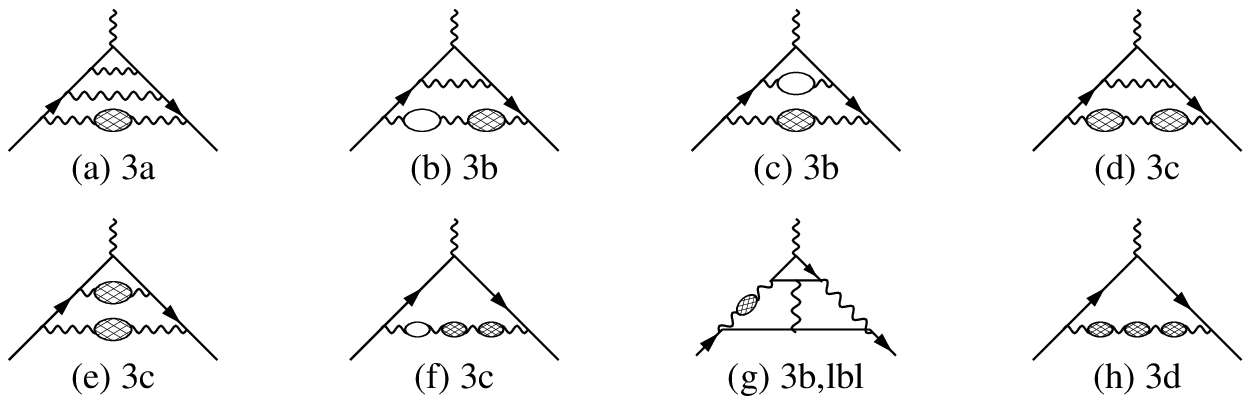}
\caption{Feynman diagrams with hadronic insertions at NLO (top row) and NNLO.}
\label{fig:ammhohad} 
\end{figure}

The next-to-leading order (NLO) HVP is represented by diagrams in
figure~\ref{fig:ammhohad}.
With kernels from~\cite{NLO}, the results of an updated evaluation are
presented in table~\ref{tab:amuNLO}.
\begin{table}[h]
\centering
\caption{NLO contributions diagrams a) - c) (in units $10^{-11}$)}
\label{tab:amuNLO}
{\small
\begin{tabular}{cccc}
\hline\noalign{\smallskip}
 $a_\mu^{(2a)}$ &$a_\mu^{(2b)}$ &$a_\mu^{(2c)}$ &$a_\mu^{{\rm had}(2)}$ \\
\noalign{\smallskip}\hline\noalign{\smallskip}
 -205.69(1.64) & 103.32(0.73)& 3.03(0.05) & -99.34 (0.91) \\ 
\noalign{\smallskip}\hline
\end{tabular}}
\caption{NNLO contributions diagrams (a) - (h) (in units $10^{-11}$)}
\label{tab:amuNNLO}
{\small
\begin{tabular}{lcrr}
\hline
Class & & Kurz et al~\cite{NNLO} & my evaluation \\
\hline
 $a_\mu^{(3a)}$ & $=$ & $\bl 8.0$ & $\bl 7.813(77)$ \\
 $a_\mu^{(3b)}$ & $=$ & $-4.1$ & $-4.023(36)$ \\
 $a_\mu^{(3b,\mathrm{lbl})}$ & $=$ &  $\bl 9.1$ & $
 \bl 8.985(77)$ \\
 $a_\mu^{(3c)}$ & $=$ &  $-0.6$ & $-0.523(5)$ \\
 $a_\mu^{(3d)}$ & $=$ &  $\bl 0.005$ & $\bl 0.00445(15)$ \\
\hline
 $a_\mu^{\mathrm{had}(3)}$ &  $=$ &  $12.4(1)$ &
 $12.26(12)$\\
\hline
\end{tabular}}
\end{table}
The next-to-next leading order (NNLO) contributions have been
calculated recently~\cite{NNLO,Steinhauser15}. Diagrams are shown in
figure~\ref{fig:ammhohad} and corresponding contributions evaluated
with kernels from~\cite{NNLO} are listed in table~\ref{tab:amuNNLO}.

The challenge for the future is to keep up with the future experiments
~\cite{Hertzog15}, which will improve the experimental accuracy from
$\delta a_\mu^\mathrm{exp} = 63 \times 10^{-11}$ [$\pm$0.54 ppm] at
present to $\delta a_\mu^\mathrm{exp}=16 \times 10^{-11}$ [$\pm$0.14
ppm] the next years. The present results $\amuh[\rm LO \ VP] =
(6873\pm35) \times 10^{-11}$ amount to +59.09 $\pm$0.30 ppm, which
poses the major challenge.  The subleading results
$\amuh [\rm NLO \ VP]= (-99.2\pm 1.0) \times 10^{-11}$ and
$\amuh [\rm NNLO \ VP] = (12.4\pm 0.1) \times 10^{-11}$ although relevant will
be known well enough. These number also compare with the well established weak
$a_\mu^{{\rm EW}}= (154\pm 1) \times 10^{-11}$ and the problematic HLbL
estimated to contribute $a_\mu^{{\rm had,LbL}}= [(105\div 106)\pm
(26\div39)]\times 10^{-11}$, which is representing a +0.90 $\pm$0.28 ppm effect.
Next generation experiments require a factor 4 reduction of the
uncertainty optimistically feasible should be a factor 2 we hope.

\section{Effective field theory: the Resonance Lagrangian Approach}
\label{sec-2}
As we know HVP is dominated by spin 1 resonance physics,
therefore we need a  low energy effective theory which includes 
\mbo{\rho,\omega,\phi} mesons. Principles to be 
implemented are the VMD mechanism, the chiral structure of QCD (chiral
perturbation theory), and electromagnetic gauge invariance. A specific
realization is the HLS effective Lagrangian~\cite{HLSOrigin}
(see~\cite{Jegerlehner:2013sja} for a brief account). In our context it
has been first applied to HLbL of muon~\mbo{g-2} in~\cite{HKS95}, to
HVP in~\cite{BDDL10}.  Largely equivalent is the ENJL model on which
the most complete analysis of HLbL in~\cite{BPP1995} was based. To
actually work in practice, the HLS symmetry has to be broken by
phenomenologically well known $S\!U(3)$ and $S\!U(2)$ flavor breaking and
the framework we consider here is the broken HLS (BHLS) model.

We briefly outline the BHLS global fit strategy of a HVP
evaluation~\cite{Maurice15}: one uses data below \mbo{E_0 = 1.05~\gv}
(just including the \mbo{\phi}) to constrain the effective Lagrangian
couplings, using 45 different data sets (6 annihilation channels and
10 partial width decays). The effective theory then can be used to
predict cross-sections mainly for two-body reactions (besides $3\pi$)
$$\ppm,~\pi^0\gamma,~\eta\gamma,~\eta'\gamma,~\pi^0\pi^+\pi^-,~K^+K^-,~K^0\bar{K}^0\,,$$
while the missing part $4\pi,5\pi,6\pi,\eta\pi\pi,\omega\pi$ as well
as the regime \mbo{E>E_0} is evaluated using data directly and pQCD
for perturbative region and tail. Including self-energy effects is
mandatory to properly describe \mbo{\rho-\omega-\phi} mixing and their
decays with proper phase space, energy dependent width etc. $\gamma-V$
($V=\rho,\omega,\phi$) mixing also turns out to be crucial. The method
works in reducing uncertainties by using indirect constraints. It is
able to reveal inconsistencies in data, e.g. KLOE vs. BaBar. A goal is
to single out a representative effective resonance Lagrangian by a
global fit, which is expected to help in improving effective field
theory (EFT) calculations of hadronic light-by-light scattering. It
has been shown~\cite{Benayoun:2011mm,Benayoun:2012wc,Benayoun:2015gxa}
that EFT not only helps reducing the uncertainty of HVP, it resolves
the $\tau$ vs. $\epm$ data puzzle, and it allows us to use, besides
the
\mbo{e^+e^-} annihilation data, also the \mbo{\tau} decay data, as
well as other experimental information consistently in a quantum field
theory framework. A best fit is obtained for the data configuration
NSK+KLOE10+KLOE12+BESIII+$\tau$ with a result
\bea
\amu^{\mathrm{had}(1)}=(682.40\pm 3.20)\times 10^{-10}\,,
\label{HLSLOHVPres}
\eea
where $(569.04 \pm 1.08)\times 10^{-10}$ results from BHLS predicted
channels and $(113.36 \pm 3.01)\times 10^{-10}$ from non-HLS [thereof
$(112.02\pm 3.01)\times 10^{-10}$ from data above 1.05 GeV and $(1.34 \pm
0.11)\times 10^{-10}$ from HLS missing channels below 1.05 GeV]. The global
fit including the BABAR sample as well yields
$(685.82\pm3.14)\times 10^{-10}$. In figure~\ref{fig:LOVPcompare} we
display the global fit BDDJ15$^*$, the best fit BDDJ15$^{\#}$ and
BDDJ12 for NSK+$\tau$, which includes scan data only.

\begin{figure}[h]
\centering
\includegraphics[width=6cm]{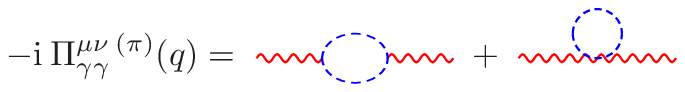}\\
\includegraphics[width=6cm]{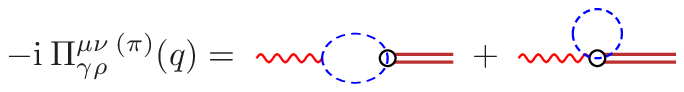}\\
\includegraphics[width=6cm]{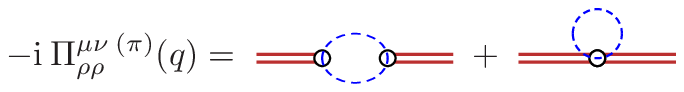}
\caption{Irreducible self-energy contribution at one-loop.}
\label{fig:rhogammSE}
\end{figure}

An important outcome of effective theory modeling of low energy hadron
physics is the observation that a unified treatment of different
processes on a Lagrangian level is able to
resolve~\cite{JS11,Maurice15} the long standing $\tau$ vs. $\epm$
$\pi\pi$ data puzzle~\cite{DEHZ03}. The main effect which distort
$\epm$-spectra relative to $\tau$-spectra is
\mbo{\rho^0-\gamma} interference in the neutral channel, which is absent
in charged channel. A minimal model, which allow us to understand
this, is VMD II plus scalar QED~\cite{JS11} for the pion-photon
interaction, with effective Lagrangian
\bea
\cL =\cL_{\gamma \rho}+\cL_\pi 
\eea
where
{\footnotesize
\bea
\hspace*{-7mm}\cL_{\gamma \rho}&=&-\frac{1}{4}\,F_{\mu\nu}\,
F^{\mu\nu}-\frac{1}{4}\,\rho_{\mu\nu}\, \rho^{\mu\nu}+
\frac{M_\rho^2}{2}\,\rho_\mu\,\rho^\mu+\frac{e}{2\,g_\rho}\,\rho_{\mu\nu}\,
F^{\mu\nu}\,, \nn \\
\hspace*{-7mm}\cL_\pi&=&D_\mu \pi^+ D^{+\mu} \pi^- -m_\pi^2 \pi^+\pi^- \semis
D_\mu = \partial_\mu -\I \,e\,A_\mu -\I\,g_{\rho\pi\pi}\,\rho_\mu\epo \nn 
\eea
}
Photon and $\rho$ self-energies are then pion-loops, which also
implies non-trivial $\gamma-\rho^0$  vacuum polarization (see figure~\ref{fig:rhogammSE}).
The clue is that the $\rho^0-\gamma$ mixing is uniquely fixed by the
electronic $\rho$-width \mbo{\Gamma_{\rho ee}}. Nothing unknown to
be adjusted! Previous calculations
\`a la Gounaris-Sakurai, considered the mixing term to be constant,
i.e. {\raisebox{-0.6ex}{\includegraphics[width=7cm]{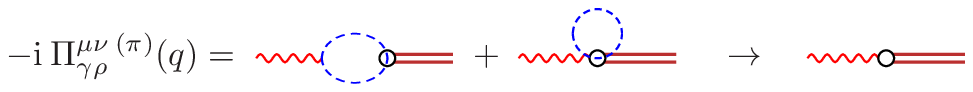}}~.  
The problem in comparing charged with neutral channel data turned out
to be an inconsistent treatment of quantum loops in the neutral channel
Gounaris-Sakurai formula. A consistent calculation requires to
consider the 2x2 matrix $\gamma-\rho$ propagator $D_{\alpha\beta}(s)$
($\alpha \in \gamma,\rho\,,\,\,\beta \in \gamma,\rho$) as a starting
point.  The off diagonal $D_{\gamma\rho}$ element usually has been
treated as the well-known VMD $\gamma-\rho$-mixing coupling constant.
However, one-loop self-energy effects should be included consistently,
especially if it turns out that interference effects are large, in
spite of the fact that one of the $g_{\rho\pi\pi}$ couplings in
$D_{\rho\rho}$ is replaced by the electromagnetic charge $e$ in
$D_{\rho\gamma}$.
\begin{figure}[h]
\centering
\includegraphics[width=0.48\textwidth]{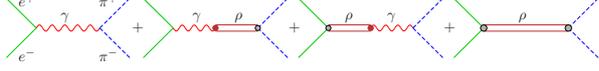}
\caption{Diagrams contributing to the process $ \eepp$. Propagators are
supposed to be corrected for self-energy effects as displayed in figure~\ref{fig:rhogammSE}.}
\label{fig:Fpipidiag}
\end{figure}
In our extended VMD model, properly renormalized, the pion form-factor
exhibits the four terms shown in figure~\ref{fig:Fpipidiag} such that $$ F_\pi(s)
\propto e^2\,D_{\gamma\gamma}+ eg_{\rho\pi\pi}\,D_{\gamma \rho}-
g_{\rho ee}e D_{\rho \gamma} -g_{\rho ee} g_{\rho\pi\pi}
\,D_{\rho\rho}\,,$$ which replaces the $\rho$ contribution of the GS
formula, which usually includes the $\omega-\rho$ mixing and higher
$\rho$ contributions $\rho'=\rho(1450)$ and $\rho''=\rho(1700)$.
Properly normalized (VP subtraction: $e^2(s)\to e^2$) we have {\small
\bea
\hspace*{-7mm}F_\pi(s) =  \left[e^2\,D_{\gamma\gamma}+ e\,(g_{\rho\pi\pi}-g_{\rho ee})\,D_{\gamma \rho}-
g_{\rho ee} g_{\rho\pi\pi} \,D_{\rho\rho}\right]/\left[e^2\,D_{\gamma\gamma}\right]\!\!\!
\label{FpipiNC}
\eea
}
Typical couplings are
$ g_{\rho\pi\pi\,\mathrm{bare}} = 5.8935$,
$ g_{\rho\pi\pi\, \mathrm{ren}} = 6.1559$, $ g_{\rho ee} =  0.018149$,
$ x=g_{\rho\pi\pi}/g_\rho=   1.15128$~\cite{JS11}.

As a result a correction displayed in figure~\ref{fig:garomixcorr}
is obtained, a +5 to -10\% correction! The proper relationship between
$\epm$ and $\tau$ spectral functions 
\bea
v_i(s)=\frac{\beta_i^3(s)}{12}\,|F^i_\pi(s)|^2~~~(i=0,-)\,,
\eea
in terms of the pion form-factors $F_\pi$ and the pion velocities $\beta_i$, now reads
\begin{figure}
\centering
\includegraphics[width=7cm]{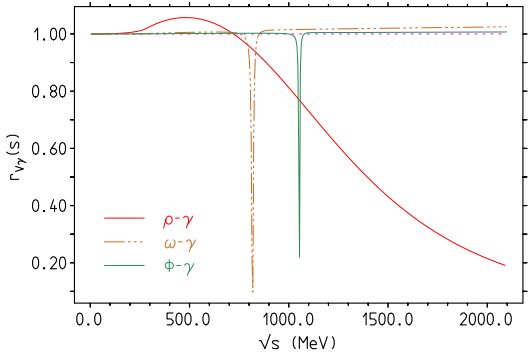}
\caption{The $\rho^0-\gamma$ mixing correction to be applied to the $\tau$
spectral functions. Corresponding effects for \mbo{\omega,\phi} in \mbo{\pi\pi} 
off-resonance are tiny (scaled up
\mbo{\Gamma_V/\Gamma(V\to\pi\pi)}. Caution: the model only applies for
energies below about $M_{\phi}$. At the $f_2(1270)$ in $\gamma \gamma \to \pi^+\pi^-,\pi^0\pi^0$
one observes the photons to couple to the quarks rather than to
point-like pions.}
\label{fig:garomixcorr}
\end{figure}
\bea
\varv_0(s)=r_{\rho\gamma}(s)\,R_{\rm IB}(s)\,\varv_-(s)\,,
\eea
where $R_{\rm IB}(s)$ is the standard isospin breaking correction
(see~\cite{GJ04,Davier:2009ag,JS11,ZhiqingZhang15}). The  
\mbo{\tau} requires to be corrected for missing
\mbo{\rho-\gamma} mixing as well, before being used as I=1 $\epm$ data, because
results obtained from \mbo{\epm} data is what goes into the DR
(\ref{amuint}) (the photon coupled to $\pi^+\pi^-$ not to
$\pi^\pm\pi^0$).  The correction is large only for the $\rho$ and
affects narrower resonances only very near resonance. The effect is
part of the experimental data and as $\omega$ and $\phi$ have no
charged partners there is nothing to be corrected in these cases. To
include further mixing effects, like $\omega-\rho^0$ mixing, one has to
extend the Lagrangian and including all possible fields and their
possible interactions, which leads to the HLS Lagrangian or a related effective
model. For details I refer to~\cite{Maurice15}. Nevertheless, let me
add a few comments on the HLS approach:

Can global fits like our HLS implementation discriminate between
incompatible data sets?  The problem of inconsistent data is not a
problem of whatever model, rather it is a matter of systematics of the
measurements. Note that modeling is indispensable for interrelating
different data channels.  In the HLS global fit \mbo{\tau} data play a
central role as they are
\textit{simple}, i.e. pure I=1, no singlet contribution, no
\mbo{\gamma-\rho^0-\omega-\phi}  mixing. In fact, \mbo{\tau}-spectra 
supplemented with PDG isospin breaking, provide a good initial fit
for most \mbo{\epem}-data fits, which then are improved and optimized
by iteration for a best simultaneous solution.
\begin{figure}[h]
\centering
\includegraphics[width=7.6cm]{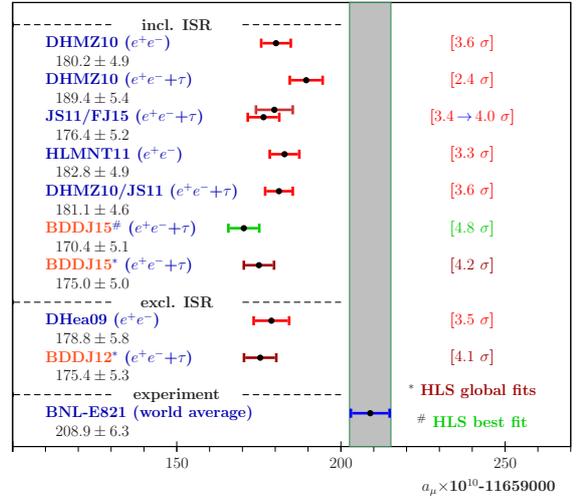}
\caption{Comparison of recent LO $\amuh$ evaluations. Note that some
results do not include \mbo{\tau} data. The HLS best fit BDDJ15{$^{\#}$}
(NSK+KLOE10+KLOE12) does not include BaBar $\pi\pi$
data~\cite{Benayoun:2015gxa}, while BDDJ15{$^*$} does.
JS11/FJ15~\cite{JS11} is updated to include the new BES III
data. Further points are BDDJ12~\cite{Benayoun:2011mm},
DHMZ10~\cite{DavierHoecker2,DavierHoecker3}, HLMNT11~\cite{Teubner2} and
DHea09~\cite{Davier:2009ag},
(see also~\cite{Davier:2015bka}).
}
\label{fig:LOVPcompare}
\end{figure}

Why do we get slightly lower results for HVP and with reduced
uncertainties? BaBar data according to~\cite{DavierHoecker2} are in
good accord with Belle \mbo{\tau}-data, \textit{before} correcting
\mbo{\tau}-data for the substantial and quite unambiguous
\mbo{\gamma-\rho^0} mixing effects! i.e. for the BaBar data alone there
seems to be no $\tau$ vs. $\epm$ puzzle, while the puzzle exists for
all other $\epm$ data sets. This is a problem for the BaBar data. They
are disfavored by our global fit! BaBar data rise the HVP estimate quite
substantially towards the uncorrected \mbo{\tau} data value. In
contrast the NSK, KLOE10/12 and the new BESIII data are in very good
agreement with the \mbo{\tau+\mathrm{PDG}}
prediction~\cite{Benayoun:2015gxa}, so they dominate the fit and give
somewhat lower HVP result\footnote{We are talking about a 1\% shift,
which is of the
order of the size of the uncertainty.}! Since, besides the
\mbo{\epem} data, additional data constrain the HLS Lagrangian and its
parameters, we find a reduced uncertainty and hence an increased
significance.

What are the model (using specifically HLS) errors of our estimates?
This is hard to say. Best do a corresponding analysis based on
different implementations of the resonance Lagrangian approach. Try to
include higher order corrections. However, the fit quality is
surprisingly good and we do not expect that one has much
flexibility. However, on can improve on photon radiation within a
suitably extended HLS approach. Such processes have been implemented
recently in the {\tt CARLOMAT} Monte Carlo~\cite{Carlomat}.

To conclude: our analysis is a first step in a direction which should allow
for systematic improvements.
A comparison of different estimates and leading uncertainties is shown
in figure~\ref{fig:LOVPcompare}.

\section{HVP for the electron anomaly}
\label{sec-3}
An up-to-date reevaluation of hadronic VP effects to the electron
$g-2$ yields the results given in table~\ref{tab:HVPae}. The present
status is illustrated by the pie chart figure~\ref{fig:gmdistc15_ele}.

\begin{table}[h]
\centering
\caption{ 2015 update of HVP effects contributing to $a_e$}
\label{tab:HVPae}
\begin{tabular}{l}
$a_e^{\mathrm{had}(1)}=(184.64\pm 1.21)\times 10^{-14}$ (LO) \\
$a_e^{\mathrm{had}(2)}=(-22.10\pm 0.14)\times 10^{-14}$ (NLO)\\
$a_e^{\mathrm{had}(3)}=(\phantom{-2}2.79\pm 0.02)\times 10^{-14}$ (NNLO)~\cite{NNLO}\\
\end{tabular}
\end{table}
\begin{figure}[h]
\centering
\includegraphics[width=7cm]{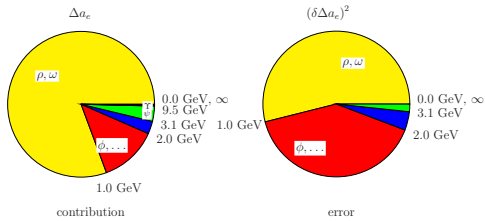}
\caption{Electron $g-2$: contributions and square errors from different energy ranges.}
\label{fig:gmdistc15_ele}
\end{figure}
On the theory side, the by far dominant QED contribution
has been calculated to 5-loops~\cite{Aoyama:2012wj}
with the result
\begin{eqnarray*}
a_e^{\rm QED}
 &=& \frac{\alpha}{2\pi}
 -0.328\,478\,444\,002\,55(33) \left( \frac{\alpha}{\pi}\right)^2
\nonumber\\&&
 +1.181\, 234\, 016\, 816(11) \left( \frac{\alpha}{\pi}\right)^3\nonumber\\
&&
-1.9097(20)\left( \frac{\alpha}{\pi}\right)^4
+9.16(58)\left(\frac{\alpha}{\pi}\right)^5.
\end{eqnarray*}
Together with the hadronic and weak contribution we get
the SM prediction (incl. $a_e^\mathrm{had, LbL}=(3.7 \pm 0.5) \power{-14}$)

\bea
a_e^{\rm SM} = a_e^{\rm QED} 
+1.720(12)\times 10^{-12}~
\mbox{(hadr \& weak)}\,,\nn
\eea
which can be used for extracting $\alpha_{\rm QED}$ from $a_e$ at
unprecedented precision. Matching the theory prediction with the very
precise experimental result of Gabrielse et al.~\cite{aenew}
\begin{eqnarray*}
a_{e}^{\rm exp}= 0.001\, 159\, 652\, 180\, 73(28)\,,
\end{eqnarray*}
one extracts
\begin{eqnarray*}
\alpha^{-1}(a_e)=137.0359991636(342)(68)(46)(24)[353]\;,
\end{eqnarray*}
which is close [36 $\to$ 57 in $10^{-10}$] to the value
\bea
\alpha^{-1}(a_e)=137.0359991657(342)[0.25\, \mathrm{ppb}]\;,
\eea
obtained by~\cite{Aoyama:2012wj}.
Note that the weak part has been reevaluated as
\bea
a_e^{\rm weak}=0.030 \times 10^{-12}\,,
\eea
which is replacing the value \mbo{0.039 \times
10^{-12}} which has been estimated in~\cite{JN2009}. An inconsistency there
has been noted by M. Passera~\cite{Passera14priv}.

The best test for new physics can be obtained by using \mbo{\alpha}
from atomic interferometry~\cite{Rb11}. With \mbo{\alpha^{-1}(\mathrm{Rb11})=137.035999037(91)[0.66\, \mathrm{ppb}]}
as an input one finds
\begin{eqnarray*}
a_{e}^{\rm the}= 0.001\, 159\, 652\, 181\, 83(77)\,,
\end{eqnarray*}
such that
\bea
a_e^\mathrm{exp}-a_e^\mathrm{the}=-1.10(0.82) \times 10^{-12}\,,
\eea
in very good agreement. We know that the sensitivity to new physics is
reduced by $(m_\mu/m_e)^2\cdot \delta a^{\rm exp}_e/\delta a^{\rm
exp}_\mu\simeq 19$ relative to $a_\mu$. Nevertheless, one has to keep
in mind that $a_e$ is suffering less form hadronic uncertainties and
thus may provide a safer test. Presently, the $a_e$ prediction is
limited by the, by a factor $\delta \alpha(\mathrm{Rb11})/\delta
\alpha(a_e)\simeq 5.3$ less precise, $\alpha$ available. Combining all
uncertainties $\amu$ is about a factor 43 more sensitive to new
physics at present.

\section{HVP subtraction of \mbo{R_\gamma(s)}: a problem of the DR method?}
\label{sec-4}
The full photon propagator is usually obtained by Dyson resummation of
the 1pi part (blob) as illustrated by figure~\ref{fig:Dyson}.
\begin{figure}[h]
\centering
\includegraphics[width=0.5\textwidth]{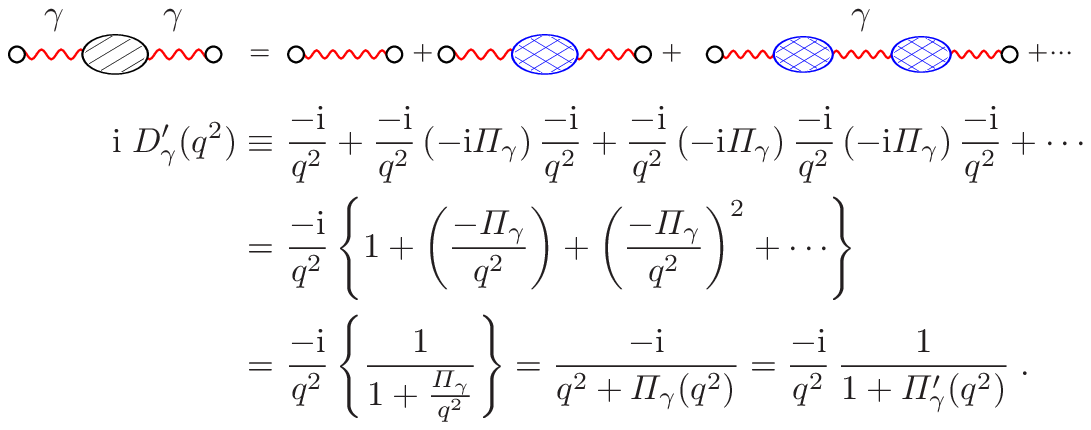}
\caption{The Dyson summation of the photon self-energy.}
\label{fig:Dyson}
\end{figure}
As we know this is a geometric series $1+x+x^2+\cdots=1/(1-x)$ which
only converges iff $|x|<1$. Including the external e.m. couplings we have 
\bea
\I\;e^2\,D'_\gamma (q^2) 
=\frac{-\I}{q^2}\,\frac{e^2}{1 + \Pi'_\gamma (q^2)}\epo
\eea
The effective charge thus is given by the well-known expression
\bea
\frac{e^2}{1 + \Pi'_\gamma (s)}=\frac{e^2}{1 - \Delta \alpha (s)}=e^2(s)\epo
\eea
Usually, \mbo{\Delta \alpha (s)} is a correction i.e \mbo{\Delta
\alpha (s) \ll 1} and the Dyson series converges well. Indeed for any
type of perturbative effects no problem is encountered (besides
possible Landau poles). For non-perturbative strong interaction
physics there are exceptions. One would expect that, if there are
problems, one would encounter them at low energy, but for the $\rho$,
the $\omega$ and the $\phi$, in spite of huge resonance enhancements,
the hadronic VP contributions to the running charge are small relative
to unity, as the effect is suppressed by the e.m. coupling $e^2$. The
exception, surprisingly, we find at pretty high energies, at the
narrow OZI suppressed resonances, which are extremely sharp, because
they lie below corresponding
\mbo{q\bar{q}}-thresholds. While the strong interaction appears
heavily suppressed (3 gluons exchange) the electromagnetic channel (1
photon exchange) appears almost as strong as the strong one (see
figure~\ref{fig:QZIsuppression}).
\begin{figure}
\centering
\includegraphics[width=5.2cm]{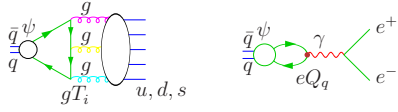}
\caption{OZI suppressed strong decays let e.m. interaction look to be
almost of equal strength.}
\vspace*{-6mm}
\label{fig:QZIsuppression}
\end{figure}
Actually, \mbo{\Gamma_{ee}} is not much smaller than \mbo{\Gamma_{\rm QCD}} (i.e
strong decays). This phenomenon shows up for the resonances
\mbo{J/\psi,\psi_2,\Upsilon_1,\Upsilon_2} and \mbo{\Upsilon_3}.
The imaginary parts from the narrow resonances read 
\bea
\Impa \Pi'_\gamma(s))=\frac{\alpha}{3}\,R_\gamma(s)=\frac{3}{\alpha}\,\frac{\Gamma_{ee}}{\Gamma}
\eea
at peak, causing the sharp spikes, which are seen only by
appropriate high resolution scans, as we know. Let $\alpha(s)$ denotes the real $\alpha(s)=\alpha/(1+\Repa \Pi'_\gamma(s))$,
we note that,
$$|1+\Pi'_\gamma(s)|^2-(\alpha/\alpha(s))^2=(\Impa \Pi'_\gamma(s))^2$$
and at \mbo{\sqrt{s}=M_R} 
values for the the different resonances are given by
$1.23 \power{-3}$ ($\rho$), $2.76 \power{-3}$ ($\omega$), $1.56
\power{-2}$ ($\phi$), $594.81$ ($J/\psi$), $9.58$ ($\psi_2$), $2.66
\power{-4}$ ($\psi_3$), $104.26$ ($\Upsilon_1$), $30.51$
($\Upsilon_2$ ) and $55.58$ ($\Upsilon_3$ ). This shows that near
QZI suppressed resonances the Dyson series cannot converge. So we have
a problem with the dispersive approach, which requires $R_\gamma(s) \propto
\Impa \Pi'_\gamma(s)$ as an input. What is measured by an experiment is the
full propagator, the summed up Dyson series, $Z=|1/(1-x)|^2$, but we
cannot extract $x$ from that since for $|x| \geq 1$ the observable $Z$
has no representation in terms of $x$.
Remember that the object required in the DR is the undressed
$R_\gamma(s)$ in~(\ref{Rfun}), 
which cannot be measured itself, rather we have to extract ($x=-\Pi'_\gamma(s)$)
\bea
R_\gamma^\mathrm{bare}=R_\gamma^{\rm phys}\,|1+\Pi'_\gamma(s)|^2\epo
\eea
Locally, near OZI suppressed resonances, the usual iterative procedure
of getting $R_\gamma^\mathrm{bare}$ does not converge! The way out
usually practiced is to utilize the smooth space-like charge, i.e. 
$\overline{R}_\gamma^\mathrm{bare}=R_\gamma^{\rm phys}\,|1+\Pi'_\gamma(-s)|^2\,,$
expected to do the undressing ``in average''. This actually does not
look too wrong as we see in figure~\ref{fig:runningalp}. Nevertheless,
I see a problem her, not only for the interpretation of resonance data,
where one would wish to be able to disentangle electromagnetic form
strong interaction effects.
\begin{figure}
\centering
\includegraphics[height=3.1cm]{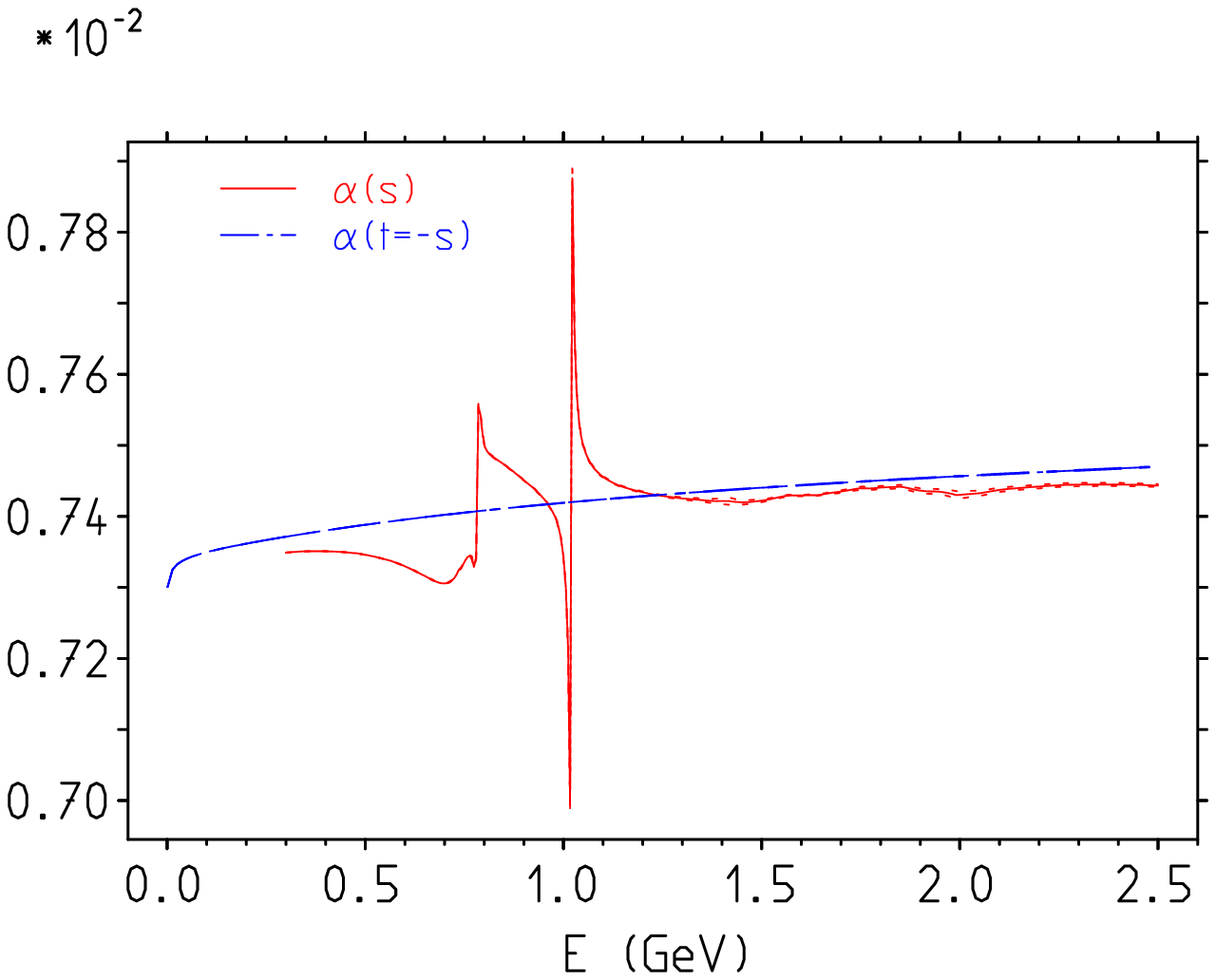}
\includegraphics[height=3.2cm]{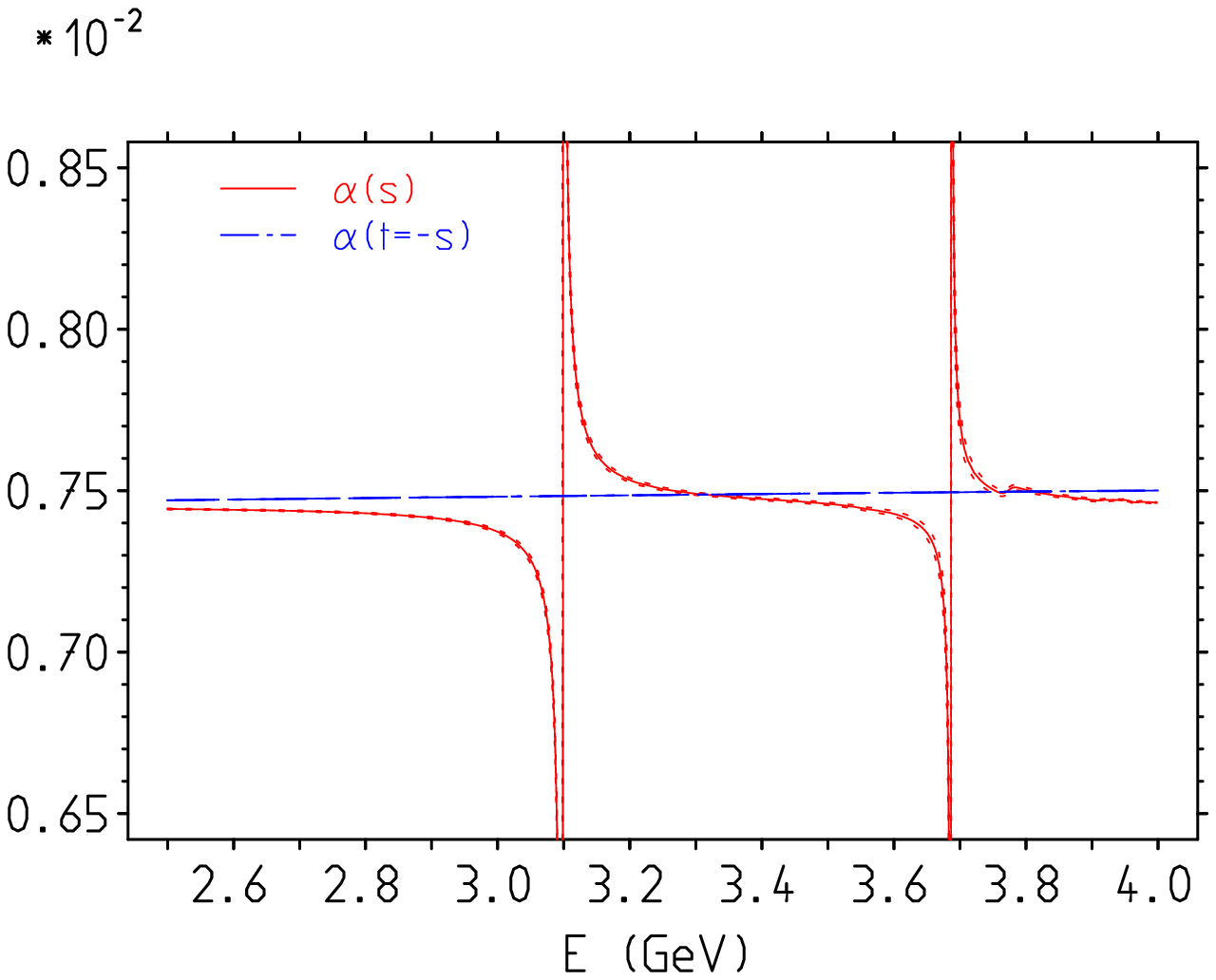}
\caption{Time-like vs. space-like effective finestructure constant $\alpha$ as a function
of the energy $E$: $\alpha(s)$ in the mean follows $\alpha(t=-s)$ ($s=E^2$).
Note that the smooth space-like effective charge agrees rather well
with the non-resonant ``background'' above the \mbo{\phi} (kind of duality).}
\vspace*{-6mm}
\label{fig:runningalp}
\end{figure}

For what concerns the proper extraction of the hadronic effects
contributing to the running of $\alpha_{\rm QED}$ and to $\amuh$, I see
no proof that this cannot produce non-negligible shifts!

Fortunately, experimental progress is in sight here: KLOE
2015~\cite{KLOEalpha} has a first direct measurements of the time-like
complex running \mbo{\alpha_{\rm QED}(s)}! Similar measurements for the
\mbo{J/\psi} and other ultra-narrow resonances should be possible with
BES III. It is a fundamental problem! An interesting possibility in
this respect is a novel approach to determine $\amuh$ form a direct
space-like measurement of $\alpha(-Q^2)$ as proposed
in~\cite{Calame:2015fva,Trentadue15},  recently.

\section{A comment on axial exchanges in HLbL}
The Landau-Yang theorem says that the amplitude $\left.\cA (\mathrm{ \
axial \  meson \ } \gamma \gamma) \right|_{\rm on-shell}=0$, 
e.g.~$ Z^0 \to \gamma \gamma$ is forbidden, while $Z^0 \to
\gamma e^+e^-$ is allowed as one of the photons is off-shell. 
For HLbL such type of contribution has been estimated in~\cite{MV2004} to
be rather large, which raised the question: Why $\amu[a_1,f_1',f_1]\;
\sim \; 22 \power{-11}$ is so large? From the data side we know, untagged
$\gamma \gamma \to f_1$ shows no signal, while single-tag $\gamma^*
\gamma \to f_1$ is strongly peaked when $Q^2 \gg m_{f_1}^2$. The
point: the contribution from axial mesons has been calculated assuming
a symmetric form-factors under exchange of the two photon
momenta. This violates the Landau-Yang theorem, which requires an
antisymmetric form-factor. In fact antisymmetrizing the form-factor
adopted in~\cite{MV2004} reduces the contribution by a factor about 3,
and the result agrees with previous findings~\cite{HKS95,BPP1995} and
with the more recent result~\cite{PaukVanderhaeghen2013}. As a result
one finds that the estimate \mbo{a_\mu^{{\rm HLbL,\,
LO}}=(116\pm39)\power{-11}} accepted in~\cite{JN2009} must be replaced
by
\bea
a_\mu^{{\rm HLbL,\, LO}}=(102\pm39)\power{-11}\epo
\eea
This also requires a modification of the result advocated
in~\cite{LBL}. The evaluation of the axialvector mesons contribution,
taking a Landau-Yang modified (i,e, antisymmetrized)
Melnikov-Vainshtein form-factors yields~\cite{FJ14} {\footnotesize
\bea
\hspace*{-6mm}\amu[a_1,f_1',f_1] \sim ({ 7.51{ =[1.89+5.04+0.58]} \pm 2.71}) \power{-11}\,,
\eea
} where ideal mixing and nonet symmetry results have been averaged. In
fact, the sum of the contributions from the $f_1$ and $f_1'$ depends
little on the mixing scheme.  The result supersedes {\footnotesize $
\amu[a_1,f_1',f_1] \sim 22(5) \power{-11}$} we included
in~\cite{JN2009}.

\section{Theory vs. experiment: do we see New Physics?}
\label{sec-5}
\begin{table}[h]
\centering
\caption{A list of small shifts in theory [in units
$10^{-11}$]. The error from new entries in
the list reduces from the old $5$ to  $3.6$.}
\label{tab:amunew}
{ \small
\begin{tabular}{lr@{$\pm$}ll}
\hline
New contribution  & \multicolumn{2}{c}{\mbo{\amu}}& Reference\\
\hline
Old axial exchange HLbL& 22&5 &\cite{MV2004} \\
New axial exchange HLbL&  7.51&2.71 &\cite{PaukVanderhaeghen2013,FJ14}\\
NNLO HVP               & 12.4 &0.1  &\cite{NNLO}\\ 
NLO HLbL               & 3     &2   &\cite{LbLNLO}\\ 
Tensor exchange HLbL   & 1.1   &0.1 &\cite{PaukVanderhaeghen2013}\\ 
\hline
Total change           &+2.0   &3.4 &
\end{tabular}
}
\end{table}
\begin{table}[t]
\centering
\caption{Standard model theory and experiment comparison [in units
$10^{-10}$].}
\label{tab:thevsexp}
{\footnotesize
\begin{tabular}{lr@{ .}lr@{ .}lc}
&\ttc{}~&\ttc{}&\\[-3mm]
\hline
&\ttc{}~&\ttc{}&\\[-3mm]
Contribution & \multicolumn{2}{c}{Value} & \multicolumn{2}{c}{Error} & Reference \\
&\ttc{}~&\ttc{}&\\[-3mm]
\hline\noalign{\smallskip}
QED at 5-loops & { 11\,658\,471}&{ 8851} & {
0}&{ 036} & \cite{Aoyama:2012wk,Kurz:2015bia,Steinhauser15}\\
LO HVP& { 687}&{ 19} & {
3}&{ 48} & (\ref{LOHVPres}) \\
NLO HVP &{ -9}&{ 934} &
{ 0}&{ 091} & table~\ref{tab:amuNLO}  \\
NNLO HVP &  { 1}&{ 226} &
{ 0}&{ 012} & ~\cite{NNLO}, table~\ref{tab:amuNNLO}  \\
HLbL &  { 10}&{ 6} & { 3}&{ 9}
& ~\cite{JN2009,FJ14}, table~\ref{tab:amunew} \\
EW at 2-loops & { 15}&{ 40} & { 0}&{ 10} & ~\cite{KPPdeR02,CMV03,Gnendiger:2013pva}  \\
&\ttc{}~&\ttc{}&\\
Theory & { 11\,659\,176}&{ 37} & { 5}&{ 18} & --  \\
Experiment & { 11\,659\,209}&{ 1} & { 6}&{ 3} & ~\cite{BNLfinal} (updated)  \\
Exp-The~ 4.0 $\sigma$  & { 32}&{ 73}
& { 8}&{ 15} & -- \\ \noalign{\smallskip}\hline
\end{tabular}}
\end{table}
Here I briefly summarize what is new and where we are. Some new
results/evaluations are collected in table~\ref{tab:amunew}.  We
finally compare the SM prediction for $\amu$ with its experimental
value~\cite{BNLfinal} in table~\ref{tab:thevsexp}, which also
summarizes the present status of the different contributions to
$\amu$.  A deviation between 3 and 5 $\sigma$ is persisting and was
slightly increasing. Resonance Lagrangian models, like the HLS model,
provide clear evidence that there is no $\tau$ version HVP which
differs from the $\epem$ data result. This consolidates a larger
deviation $\Delta a_\mu=a_\mu^{\rm exp}-a_\mu^{\rm the}$. Also the
decrease of the axial HLbL contribution goes in this direction, it is
compensated however by the new NNLO HVP result.  What represents the 4
$\sigma$ deviation: new physics?  Is it a statistical fluctuation? Are
we underestimating uncertainties (experimental, theoretical)? Do
experiments measure what theoreticians calculate? I refer
to~\cite{Stoeckinger15} for possible interpretations and conclusions.

\section{Outlook}
\label{sec-6}
Although progress is slow, there is evident progress in reducing the
hadronic uncertainties, most directly by progress in measuring the
relevant hadronic cross-sections. Near future progress we expect from
BINP Novosibirsk/Russia and from IHEP Beijing/China. Energy scan as
well as ISR measurement of cross-sections in the region from 1.4 to
2.5 GeV are most important to reduce the errors to a level competitive
with the factor 4 improvement achievable by the upcoming new muon
$g-2$ experiments at Fermilab/USA and at JPAC/Japan~\cite{Hertzog15}.
Also BaBar data are still being analyzed and are important for
improving the results. Promising is that lattice QCD evaluations come
closer to be competitive~\cite{Petschlies15}.

\acknowledgement
I thank Z. Zhang for helpful discussions on the isospin breaking corrections
and M. Benayoun for close collaboration on HLS driven estimates of $\amuh$.

\vspace*{-1cm}

\begin{thebibliography}{}

\bibitem{HKS95}
M.~Hayakawa, T.~Kinoshita, A.~I.~Sanda,
Phys.\ Rev.\ Lett.\  {\bf 75}, 790 (1995);
Phys.\ Rev.\ D {\bf 54}, 3137 (1996)

\bibitem{BPP1995}
J.~Bijnens, E.~Pallante, J.~Prades,
Phys.\ Rev.\ Lett.\  {\bf 75}, 1447 (1995)
[Erratum-ibid.\  {\bf 75}, 3781 (1995)];
Nucl.\ Phys.\ B {\bf 474}, 379 (1996);
[Erratum-ibid.\ {\bf 626}, 410 (2002)];\\
J.~Bijnens, J.~Prades,
Mod.\ Phys.\ Lett.\  A {\bf 22}, 767 (2007)

\bibitem{Bijnens15}
  J.~Bijnens,
  arXiv:1510.05796 [hep-ph];
these proceedings

\bibitem{EJ95}
S.~Eidelman, F.~Jegerlehner,
Z.\ Phys.\ C {\bf 67}, 585 (1995);
F.~Jegerlehner,
Nucl.\ Phys.\ (Proc.\ Suppl.) C  {\bf 51}, 131 (1996);
J.\ Phys.\ G {\bf 29}, 101 (2003);
Nucl.\ Phys.\ Proc.\ Suppl.\  {\bf 126}, 325 (2004)

\bibitem{Benayoun:2011mm}
M.~Benayoun, P.~David, L.~DelBuono, F.~Jegerlehner,
  Eur.\ Phys.\ J.\ C {\bf 72}, 1848 (2012)

\bibitem{Benayoun:2012wc}
  M.~Benayoun, P.~David, L.~DelBuono, F.~Jegerlehner,
  Eur.\ Phys.\ J.\ C {\bf 73}, 2453  (2013)

\bibitem{Benayoun:2015gxa}
  M.~Benayoun, P.~David, L.~DelBuono, F.~Jegerlehner,
  arXiv:1507.02943 [hep-ph]

\bibitem{Jegerlehner:2013sja}
  F.~Jegerlehner,
  Acta Phys.\ Polon.\ B {\bf 44}, Vol.11, 2257 (2013)

\bibitem{Maurice15}
  M.~Benayoun,
  arXiv:1511.01329 [hep-ph];
these proceedings.

\bibitem{Boyle:2011hu}
  P.~Boyle, L.~Del Debbio, E.~Kerrane, J.~Zanotti,
  Phys.\ Rev.\ D {\bf 85} (2012) 074504

\bibitem{Feng:2013xsa}
  X.~Feng et al.,
  Phys.\ Rev.\ D {\bf 88}, 034505 (2013)

\bibitem{Aubin:2013daa}
  C.~Aubin, T.~Blum, M.~Golterman, S.~Peris,
  Phys.\ Rev.\ D {\bf 88}, 074505 (2013) 

\bibitem{Francis:2014dta}
  A.~Francis et al.,
  arXiv:1411.3031 [hep-lat]

\bibitem{Malak:2015sla}
  R.~Malak et al. [Budapest-Marseille-Wuppertal Collab.],
  PoS LATTICE {\bf 2014}, 161 (2015)

\bibitem{Petschlies15}
M.~Petschlies, 
these proceedings.

\bibitem{KnechtNyffeler01}
M.~Knecht, A.~Nyffeler,
Phys.\ Rev.\ D {\bf 65}, 073034 (2002)

\bibitem{MV2004}
K.~Melnikov, A.~Vainshtein,
Phys.\ Rev.\ D {\bf 70}, 113006 (2004)

\bibitem{Knecht15}
  M.~Knecht,
  Nucl.\ Part.\ Phys.\ Proc.\  {\bf 258-259}, 235 (2015);
and these proceedings

\bibitem{Nyffeler15}
A.~Nyffeler, 
these proceedings

\bibitem{PaukVanderhaeghen2013}
  V.~Pauk, M.~Vanderhaeghen,
  Eur.\ Phys.\ J.\ C {\bf 74}, 3008 (2014)

\bibitem{Colangelo:2014pva}
  G.~Colangelo et al.,
  Phys.\ Lett.\ B {\bf 738}, 6 (2014)

\bibitem{Procura15}
M.~Procura, 
these proceedings

\bibitem{Blum:2014oka}
  T.~Blum, S.~Chowdhury, M.~Hayakawa, T.~Izubuchi,
  Phys.\ Rev.\ Lett.\  {\bf 114}, 012001 (2015)

\bibitem{Lehner15}
Ch.~Lehner, 
these proceedings

\bibitem{KPPdeR02}
M.~Knecht, S.~Peris, M.~Perrottet, E.~de Rafael, 
JHEP {\bf 0211}, 003 (2002)

\bibitem{CMV03}
A.~Czarnecki, W.~J.~Marciano, A.~Vainshtein,
Phys.\ Rev.\ D {\bf 67}, 073006 (2003) [Erratum-ibid.\  D {\bf 73}, 119901 (2006)] 


\bibitem{CMD203}
R.~R.~Akhmetshin et al.  [CMD-2 Collab.],
Phys.\ Lett.\ B {\bf 578}, 285 (2004)

\bibitem{CMD206}
V.~M.~Aulchenko et al.  [CMD-2 Collab.],
JETP Lett.\  {\bf 82}, 743 (2005)
[Pisma Zh.\ Eksp.\ Teor.\ Fiz.\  {\bf 82}, 841 (2005)];
R.~R.~Akhmetshin et al.,
JETP Lett.\  {\bf 84}, 413 (2006)
[Pisma Zh.\ Eksp.\ Teor.\ Fiz.\  {\bf 84}, 491 (2006)];
Phys.\ Lett.\  B {\bf 648}, 28 (2007)

\bibitem{SND06}
  M.~N.~Achasov et al. [SND Collab.],
  J.\ Exp.\ Theor.\ Phys.\  {\bf 103}, 380 (2006)
   [Zh.\ Eksp.\ Teor.\ Fiz.\  {\bf 130}, 437 (2006)]

\bibitem{KLOE08}
  A.~Aloisio et al. [KLOE Collab.],
  Phys.\ Lett.\  B {\bf 606}, 12 (2005);\\
  F.~Ambrosino et al. [KLOE Collab.],
  Phys.\ Lett.\  B {\bf 670}, 285 (2009)

\bibitem{KLOE10}
  F.~Ambrosino et al.  [KLOE Collab.],
  Phys.\ Lett.\ B {\bf 700}, 102 (2011)

\bibitem{KLOE12}
D.~Babusci et~al. [KLOE Collab.],
Phys.Lett. {\bf B720}, 336 (2013)


\bibitem{BABARpipi}
  B.~Aubert et al.  [BABAR Collab.],
  Phys.\ Rev.\ Lett.\  {\bf 103}, 231801 (2009);
  J.~P.~Lees et al.,
  Phys.Rev. {\bf D86}, 032013 (2012)

\bibitem{BESIII}
  M.~Ablikim et al. [BESIII Collab.],
  arXiv:1507.08188v3 [hep-ex]

\bibitem{JS11}
  F.~Jegerlehner, R.~Szafron,
  Eur.\ Phys.\ J.\ C {\bf 71}, 1632 (2011)


\bibitem{ADH98}
R.~Alemany, M.~Davier, A.~H\"ocker,
Eur.\ Phys.\ J.\ C {\bf 2}, 123 (1998)

\bibitem{DEHZ03}
M.~Davier, S.~Eidelman, A.~H\"ocker, Z.~Zhang,
Eur.\ Phys.\ J.\ C {\bf 27}, 497 (2003);
Eur.\ Phys.\ J.\ C {\bf 31}, 503 (2003)

\bibitem{GJ04}
  S.~Ghozzi, F.~Jegerlehner,
  Phys.\ Lett.\ B {\bf 583}, 222 (2004)

\bibitem{Davier:2009ag}
  M.~Davier et al.,
  Eur.\ Phys.\ J.\  C {\bf 66}, 127 (2010)

\bibitem{ZhiqingZhang15}
  Z.~Zhang,
  arXiv:1511.05405 [hep-ph], these proceedings

\bibitem{ALEPH}
R.~Barate et al. [ALEPH Collab.],
Z.\ Phys.\ C {\bf 76}, 15 (1997);
Eur.\ Phys.\ J.\ C {\bf 4}, 409 (1998);
S.~Schael et al. [ALEPH Collab.],
Phys.\ Rept.\  {\bf 421}, 191 (2005)

\bibitem{AlephCorr}
M.~Davier et al.,
Eur.\ Phys.\ J.\ {\bf C74}, 2803 (2014)

\bibitem{OPAL}
K.~Ackerstaff et al. [OPAL Collab.],
Eur.\ Phys.\ J.\ C {\bf 7}, 571 (1999)

\bibitem{CLEO}
S.~Anderson et al. [CLEO Collab.],
Phys.\ Rev.\ D {\bf 61}, 112002 (2000)

\bibitem{Belle}
  M.~Fujikawa et al.  [Belle Collab.],
  Phys.\ Rev.\ D {\bf 78}, 072006 (2008)

\bibitem{Akhmetshin:2013xc}
R.~Akhmetshin et~al. [CMD-3 Collab.],
Phys.Lett. {\bf B723}, 82 (2013)

\bibitem{Achasov:2013btb}
M.~Achasov et~al. [SND Collab.],
Phys.Rev. {\bf D88}, 054013 (2013)

\bibitem{Lees:2013ebn}
J.~Lees et~al. [BABAR Collab.],
Phys.Rev. {\bf D87}, 092005 (2013)

\bibitem{Lees:2013gzt}
J.~Lees et~al. [BABAR Collab.],
Phys.Rev. {\bf D88}, 032013 (2013)

\bibitem{Lees:2014xsh}
J.~Lees et~al. [BABAR Collab.],
Phys.Rev. {\bf D89}, 092002 (2014)

\bibitem{Davier:2015bka}
M.~Davier,
Nucl.\ Part.\ Phys.\ Proc.\ {\bf 260}, 102 (2015)

\bibitem{BES02}
J.~Z.~Bai et al.  [BES Collab.],
Phys.\ Rev.\ Lett.\  {\bf 84}, 594 (2000);
Phys.\ Rev.\ Lett.\  {\bf 88}, 101802 (2002);\\
  M.~Ablikim et al.,
  Phys.\ Lett.\ B {\bf 677}, 239 (2009)

\bibitem{NLO}
B.~Krause,
Phys.\ Lett.\ B {\bf 390} (1997) 392 

\bibitem{NNLO}
A.~Kurz, T.~Liu, P.~Marquard, M.~Steinhauser,
Phys.Lett. {\bf B734}, 144 (2014)

\bibitem{Steinhauser15}
M.~Steinhauser,
these proceedings

\bibitem{Hertzog15}
D.~Hertzog, 
these proceedings.

\bibitem{HLSOrigin}
M.~Bando, T.~Kugo, K.~Yamawaki,
Phys. Rept. {\bf 164}, 217 (1988);\\
M.~Harada, K.~Yamawaki,
Phys. Rept. {\bf 381}, 1 (2003)

\bibitem{BDDL10}
M.~Benayoun et al.,
Eur. Phys. J. {\bf C55}, 199 (2008);
M.~Benayoun, P.~David, L.~DelBuono, O.~Leitner,
Eur. Phys. J. {\bf C65}, 211 (2010);
%
Eur. Phys. J. {\bf C68}, 355 (2010)

\bibitem{DavierHoecker2}
M.~Davier et al.,
Eur. Phys. J. {\bf C66}, 1 (2009)

\bibitem{DavierHoecker3}
M.~Davier et al.,
Eur. Phys. J. {\bf C71}, 1515 (2011)

\bibitem{Teubner2}
K.~Hagiwara et al.,
J.~Phys. {\bf G38}, 085003 (2011)


\bibitem{Carlomat}
  K.~Ko\l odziej,
  Comput.\ Phys.\ Commun.\  {\bf 196}, 563 (2015)

\bibitem{Aoyama:2012wj}
  T.~Aoyama, M.~Hayakawa, T.~Kinoshita, M.~Nio,
  Phys.\ Rev.\ Lett.\  {\bf 109}, 111807 (2012)

\bibitem{aenew} 
G.~Gabrielse et al.,
Phys.\ Rev.\ Lett.\  {\bf 97} (2006) 030802
[Erratum-ibid.\  {\bf 99} (2007) 039902]


\bibitem{JN2009}
  F.~Jegerlehner, A.~Nyffeler,
  Phys.\ Rept.\  {\bf 477}, 1 (2009)

\bibitem{Passera14priv}
M. Passera private communication


\bibitem{Rb11}
  R.~Bouchendira et al.,
  Phys.\ Rev.\ Lett.\  {\bf 106} (2011) 080801

\bibitem{KLOEalpha}
The KLOE Collaboration, to be published
 
\bibitem{Calame:2015fva} 
  C.~M.~Carloni et al.,
  Phys.\ Lett.\ B {\bf 746}, 325 (2015)

\bibitem{Trentadue15}
L.~Trentadue, these proceedings


\bibitem{LBL}
  J.~Prades, E.~de Rafael, A.~Vainshtein,
  Adv.\ Ser.\ Direct.\ High Energy Phys.\  {\bf 20}, 303 (2009)

\bibitem{FJ14}
F. Jegerlehner, Talk at the MITP Workshop 
``Hadronic contributions to the muon anomalous magnetic moment'',
1-5 April 2014, Waldthausen Castle near Mainz,
and to be published

\bibitem{LbLNLO}
G.~Colangelo et al.,
Phys.Lett. {\bf B735}, 90 (2014)

\bibitem{Aoyama:2012wk}
  T.~Aoyama, M.~Hayakawa, T.~Kinoshita, M.~Nio,
  Phys.\ Rev.\ Lett.\  {\bf 109}, 111808 (2012)

\bibitem{Kurz:2015bia} 
  A.~Kurz et al.,
  Phys.\ Rev.\ D {\bf 92}, 073019 (2015)

\bibitem{Gnendiger:2013pva}
  C.~Gnendiger, D.~St\"ockinger, H.~St\"ockinger-Kim,
  Phys.\ Rev.\ D {\bf 88}, 053005 (2013)

\bibitem{BNLfinal}
G.~W.~Bennett et al.  [Muon g-2 Collab.],
Phys.\ Rev.\ D {\bf 73} (2006) 072003

\bibitem{Stoeckinger15}
D.~St\"ockinger, 
these proceedings
\end{thebibliography}
\end{document}